\documentstyle[a4,oldlfont,amsfonts]{article}

\makeatletter
\newmathalphabet*\bbl{msb}{m}{n}

\newmathalphabet*\eck{eur}{m}{n}

\newmathalphabet*\sno{eus}{m}{n}

\newmathalphabet*\got{euf}{m}{n}
\makeatother

\def\nn{\nonumber}
\def\non{\nonumber\\}
\def\be{\begin{equation}}
\def\ee{\end{equation}}
\def\ben{\begin{displaymath}}
\def\een{\end{displaymath}}
\def\ba{\begin{eqnarray}}
\def\ea{\end{eqnarray}}

\def\3{\ss}
\def\a{\alpha}
\def\b{\beta}
\def\c{\gamma}

\def\D{\Delta}
\def\d{\delta}

\def\f{\varphi}

\def\k{\kappa}
\def\l{\lambda}
\def\L{\Lambda}

\def\n{\eta}
\def\p{\pi}
\def\P{\Pi}

\def\r{\rho}
\def\s{\sigma}

\def\th{\theta}
\def\Th{\Theta}

\def\W{\Omega}

\def\pol{\frac{1}{z-w}\;}
\def\2pol{\frac{1}{(z-w)^2}\;}
\def\halb{\textstyle{\frac{1}{2}}}

\def\H{{\cal H}}
\def\O{{\cal O}}

\newcommand{\resection}[1]{\setcounter{equation}{0}\section{#1}}

\begin{document}
%
%%%%%%%%% Title %%%%%%%%%%%%%%%%%%%%%%%%%%%%%%%%%%%%%%%%%%%%%%%%%%%%%%%%%%
%
\thispagestyle{empty}
\renewcommand{\thefootnote}{\fnsymbol{footnote}}
\begin{flushright} hep-th/9505143 \end{flushright}
\vspace*{17mm}
\centerline{\LARGE Quantum version of the N=8 superconformal algebra}
 \vspace*{8mm}
\begin{center}
       {\sc J.\,A.\,Henning Samtleben \footnote{e-mail: {
jahsamt@x4u2.desy.de}}}\\
 \vspace*{2mm}
     IInd Institute for Theoretical Physics, University of Hamburg\\
     Luruper Chaussee 149, 22761 Hamburg, Germany\\
\vspace*{6mm} May, 1995
\end{center}
\vspace*{2cm}
\begin{abstract}\noindent
The quantized form of the soft N=8 superconformal algebra is investigated. Its
operator product expansions are shown to exhibit a one-parameter-class of
(soft) anomalies, which may be arbitrarily shifted by certain suitable quantum
corrections of the generators. In particular, the BRST operator can be
constructed and made nilpotent in the quantum version of all known realizations
of the algebra. This generalizes the results of Cederwall and Preitschopf, who
studied the $\widehat{S^7}$-algebra, that is contained as a soft Kac-Moody part
in the superconformal algebra. A Fock-space representation is given, that has
to be somewhat unusual in certain modes.
\end{abstract}
\renewcommand{\thefootnote}{\arabic{footnote}}
\setcounter{footnote}{0}
\newpage

%%%%%%%%%%%%%%%%%%%%%%%%%%%%%%%%%%%%%%%%%%%%%%%%%%%%%%%%%%%%%%%%%%%%%%%%%%
%   Text
%%%%%%%%%%%%%%%%%%%%%%%%%%%%%%%%%%%%%%%%%%%%%%%%%%%%%%%%%%%%%%%%%%%%%%%%%%
\resection{Introduction}
There has been broad interest in conformal and superconformal symmetries
throughout the last years, not at least, because of the fundamental role, they
seem to play in string theory. In order to gain insight into what might be the
underlying mathematical structures of string theory, it has even become
customary to study the possible superconformal symmetry structures for
themselves before investigating concrete models, thereby trying to rule out
inconsistent theories by looking for algebraic inconsistencies, such as
inadmissible OPE or BRST-anomalies.\\
Since the work of Ademollo et al. \cite{ademollo}, extended superconformal
algebras containing ${\rm N}>1$ super-generators and additional inner
symmetries have been under closer investigation, generalizing the first
superconformal structures, that appeared as symmetry algebras of the N=1
superstring \cite{rsstring}.  Of further relevance have been the discovered N=2
and N=4 superconformal Lie-algebras, that contain $\widehat{{\got u}(1)}$ and
$\widehat{{\got su}(2)}$ Kac-Moody parts, respectively, and can be interpreted
in the context of the corresponding superstring models, though the physical
contents of these models has remained unclear \cite{ademollo2,ademollo4}. \\
There is a close connection between these algebras and the corresponding
division algebras ${\bbl K}_{\rm N}$ of the real, the complex and the
quaternionic numbers, that becomes manifest, for example, in rewriting their
Kac-Moody parts as spheres $\widehat{S^{{\rm N}-1}}$.\\
Extending this construction to the algebra of octonions ${\bbl O}$, one arrives
at a superconformal algebra with N=8 super-generators and additional
$\widehat{S^7}$ part. As ${\bbl O}$ fails to be associative, the resulting
superconformal algebra cannot be a Lie-algebra, which is in accordance with the
general results on classification of superconformal algebras \cite{class}. The
only known way to implement non-associative algebras in physical context is via
so-called soft algebras \cite{sohnius}, i.e. via algebras with field-dependent
structure ``constants'', that transform under the algebra, thereby fulfilling
the Jacobi-identities, which seems to be essential for any physical use.\\
The N=8 superconformal algebra was discovered by Englert et al. \cite{englert}
and reappeared in physical context in the work of Berkovits \cite{berk} and
Brink, Cederwall and Preitschopf \cite{brcedpr} in different connections with
the ten-dimensional N=1 GS-superstring. In both approaches, the known
superconformal Lie-algebras, that describe the situation in the lower
dimensions $D={\rm N}+2=3,4,6$,~ are replaced by the N=8 algebra in the
critical and therefore most interesting dimension of $D=10$.\\
\\
Up to now, only the classical form of this algebra has been studied; in
particular, any terms of order $\hbar^2$ and, correspondingly, all
normal-ordering problems have been neglected. Cederwall and Preitschopf have
investigated the soft $\widehat{S^7}$ algebra in a systematic way, given
several realizations of this structure, calculated the OPE of the quantized
algebra and described a nilpotent BRST-operator \cite{cedpr}. In the following,
we will extend parts of these investigations onto the whole soft N=8
superconformal algebra.
The paper is organized as follows: In section \ref{bn} basic facts on the
alternative division algebra of the octonions ${\bbl O}$ are repeated; section
\ref{cf} describes the classical form of the N=8 superconformal algebra. We
investigate the quantized form of the soft algebra in section \ref{qf},
calculating all the arising anomalies, that do not restrict to central
extensions as in the case of the superconformal Lie-algebras, but contain
several soft terms from derivations of the structure ``constants''. It is shown
how to correct an essential part of these anomalies by quantum corrections of
the generators and arising normal-ordering problems are discussed.\\
In section \ref{gf} we give a Fock-space representation of the soft algebra,
which exhibits some unusual features due to the existence of a certain inverse
operator, that is necessary for the parameterization of the underlying
seven-sphere $S^7$. The BRST-operator is constructed and its quantum nilpotency
shown to be achievable in all known representations of the algebra by a
suitable quantum correction of the discovered type. This generalizes the
results of \cite{cedpr}.\\

\resection{Basic notation}\label{bn}
In this section we will give a short review of the known facts about the
algebra of octonions and clarify our notation, following \cite{cedpr}.\\
\\
The algebra of octonions is the only real finite-dimensional non-associative
but alternative division algebra. Let its generators be $e^0=1,e^1,\dots,e^7$
and denote the involution of an octonion $\l=\l^0 + \sum_{i=1}^{7} \l^ie^i$ by
$\l^*:=\l^0 - \sum_{i=1}^{7} \l^ie^i$. Its real and imaginary part are given by
\ben
[\l] := \halb(\l + \l^*) \qquad\quad \{\l\} := \halb(\l - \l^*)
\een
The multiplication of imaginary octonions can be defined as follows:
\be
e^ie^j = -\d^{ij}+\s^{ijk}e^k \hspace{30mm} i,j \in \{1,2,\dots,7\}
\label{mult}
\ee
with
\ben
\s^{ijk} = 1 {\rm { \hspace{3mm} for\hspace{3mm}} } [ijk] =
[124],[235],[346],[457],[561],[672],[713]~;
\een
antisymmetric in all three indices and zero otherwise \footnote{See Dixon
\cite{dixon} for all possibilities to define this multiplication.}.\\
This means, for $a,b,c \in \{0,1,\dots,7\} $:
\be
e^ae^b = \D^{abc}e^c {\rm \hspace{5mm}with\hspace{5mm}} \D^{abc} =
\d^{a0}\d^{bc} +\d^{b0}\d^{ac} - \d^{c0}\d^{ab} + \s^{abc}\label{delta}
\ee
In addition to the commutator
\be
[e^i,e^j] = 2\s^{ijk}e^k \label{comm}
\ee
one can introduce an associator measuring the deviation of associativity:
\be
[e^i,e^j,e^k] := (e^ie^j)e^k-e^i(e^je^k) =: 2\r^{ijkl}e^l\nn
\ee
An algebra is called {\em alternative}\, if the associator is antisymmetric in
its three arguments. In the case of the octonions this can be verified using
(\ref{mult}).\\
\\
Due to non-associativity several octonionic products can be introduced; the
following will be of further importance:\\
Let $X\in S^7$ be a unit octonion. The so-called {\em $X$-product} may be
defined as
\be
A\circ_X B := (AX^*)(XB) = X^*((XA)B) = (A(BX^*))X
\ee
Now {\em $X$-commutator} and {\em $X$-associator} are defined in analogy to
above:
\ba
[e^i,e^j]_X &:=& e^i\circ_X e^j -e^j\circ_X e^i \hspace{25.5mm} ~=:~
2T^{ijk}(X)~e^k \label{xkomm&ass}\\
\,[e^i,e^j,e^k]_X &:=& (e^i\circ_X e^j)\circ_X e^k - e^i\circ_X (e^j\circ_X
e^k) ~=:~ 2R^{ijkl}(X)~e^l\nn
\ea
with
\ba
R^{ijkl}(X) &=& \d^{jk}\d^{il} - \d^{jl}\d^{ik} + T^{mij}(X)T^{klm}(X)
\label{xassociator}\\
          &=& T^{m[ij}(X)T^{k]lm}(X) \nn
\ea
Note that $R^{ijkl}(X)$ is even antisymmetric in all four
indices\footnote{Throughout the whole text, symmetrization (\dots) and
anti-symmetrization [\dots] is understood with total weight one, e.g.
$[ab]=\halb(ab-ba)$.} and that $X$-commuta\-tor and $X$-associator are purely
imaginary.\\
\\
The quantities $T^{ijk}(X)$ and $R^{ijkl}(X)$ characterizing the $X$-product
are functions on the seven-sphere that may be interpreted geometrically as the
torsion and its covariant derivative \cite{englert} with
\ba
T^{ijk}(1) &=& T^{ijk}(-1) ~ =~ \s^{ijk} \label{stelle1}\\
R^{ijkl}(1) &=& R^{ijkl}(-1) ~ = ~ \r^{ijkl}\nn
\ea
Nearly all relations between these quantities can be derived from
(\ref{xassociator}) and the following two important identities:
\ba
T^{mi(a}T^{b)mj} &=& \d^{ab}\d^{ij}-\d^{i(a}\d^{b)j} \label{i1}\\
T^{abm}R^{mijk} &=&  3\bigg( \d^{a[i}T^{jk]b} -
\d^{b[i}T^{jk]a}\bigg),\label{i2}
\ea
which can be proved using (\ref{xassociator}). \\
\\
Let us close this section by stating some octonionic identities that will
become important in further calculations:
\ba
T^{imn}T^{jmn} &=& 6\d^{ij}\label{oi}\\
T^{mnk}R^{mnij} &=& 4T^{ijk}\\
R^{mnab}R^{mncd} &=& -2R^{abcd}+4T^{abm}T^{mcd}\\
&=& 2R^{abcd} + 4\d^{ac}\d^{bd} - 4\d^{ad}\d^{bc}\\
R^{maij}T^{kbn}+R^{mbij}T^{kan}
&=&R^{ma[ij}T^{k]bn}+R^{mb[ij}T^{k]an}\label{smouf}\\
2R^{mijk}T^{abm}&=&-3R^{mab[i}T^{jk]m}\label{gut}
\ea\smallskip

\resection{Classical version of the N=8 superconformal algebra}\label{cf}

In this section we will present the classical form of the soft N=8
superconformal algebra. This algebra was discovered by Englert et al. in its
non-associative form and in a soft form that turned out to be somewhat
unwieldy, not admitting central extensions for example \cite{englert}. With a
slight modification concerning the nature of the field dependent structure
``constants'', the soft algebra reappeared in the work of Berkovits \cite{berk}
and Brink, Cederwall and Preitschopf \cite{brcedpr}. They even found free field
realizations of this algebra which we are going to describe in the following.\\
\paragraph{Parameter field realization}
The simplest realization of the N=8 superconformal algebra is built from
octonionic conjugate bosonic fields $\l^a, \p^a_{(\l)}$ and conjugate fermionic
fields $\th^a, \p^a_{(\th)}$ with fundamental correlations
\be
\l^a(z)\p^b_{(\l)}(w) ~\sim~ \frac{\d^{ab}}{z-w}~ \sim
{}~\th^a(z)\p^b_{(\th)}(w)\label{ff}
\ee
The generators of the algebra are given by
\ba
  j &=& \{\p_{(\l)}^*\l\} \label{parant} \\
  g &=& \p_{(\th)}^*\l - \partial\th^*\p_{(\l)} \non
  l &=& \halb:[\l^*\partial\p_{(\l)} - \partial \l^*\p_{(\l)}]:
           - :[\p_{(\th)}^*\partial\th]:\nn
\ea
in octonionic notion $j:=j^ie_i$,~ $g:=g^ae_a$. The classical form of their OPE
is
\ba
 J^i(z)\,J^j(w) &\sim& \frac{2}{z-w}\;T^{ijk}(X(w))\,J^k(w)\label{OPEkl}\\
 J^i(z)\,G^a(w) &\sim& \frac{1}{z-w}\;\D^{bia}(X(w))\,G^b(w) +
                     \frac{2}{z-w}\;R^{airj}(X(w))\,\n^r(w)\,J^j(w)\non
 G^a(z)\,G^b(w) &\sim& \frac{2}{(z-w)^2}\;\D^{aib}(X(w))\,J^i(w) +
                    \frac{1}{z-w}\;\partial_w(\D^{aib}(X(w))\,J^i(w)) \non
&&{}+ \frac{2\d^{ab}}{z-w}\,L(w)\non
 L(z)\,\O (w) &\sim& \frac{h_{\O}}{(z-w)^2}\;\O(w) +
                  \frac{1}{z-w}\;\partial\O (w)~,\nn
\ea
with
\be
X=\frac{\l}{|\l|}\quad {\rm and}\quad \n = \frac{1}{|\l|}\;\partial\th^*X
\label{conn}
\ee
and neglecting all terms of order $\hbar^2$. The conformal dimensions of the
currents are $h_J=1$ and $h_G=\frac32$.\\
\\
These are the OPE of a {\em soft} algebra, which means that the structure
``constants'' on the r.h.s. still depend on some fields and can be interpreted
as functions on an underlying manifold, which is the seven-sphere $S^7$ and its
super-partner in this case, parameterized by $X$ and $\n$. As these fields
transform under the algebra, the structure ``constants'' have singular OPE with
the currents of the algebra:
\ba
j^i(z)\,X(w) &\sim& \pol X(w)e^i\\
j^i(z)\,\n(w) &\sim& \pol \n(w)\circ_{X(w)}e^i\quad{\rm , etc.}\nn
\ea
The resulting additional terms are essential for the validity of the Jacobi
identities.\\
\\
These soft OPE generalize the OPE of the known superconformal Lie-algebras for
N=1, N=2 and N=4 \cite{rsstring, ademollo2, ademollo4}, that may be recovered
by replacing the octonions by the corresponding algebra ${\bbl K}_{\rm N}$ of
the real, the complex or the quaternionic numbers, respectively.\\
In order to stress the division algebra structure of the OPE (\ref{OPEkl}), we
still present them in the following form:
\ba
 J^i(z)\,J(w) &\sim& \frac{1}{z-w}\;[J(w),e^i\,]_{X(w)} \label{OPEokt}\\
 J^i(z)\,G(w) &\sim& \frac{1}{z-w}\;G(w)\circ_{X(w)}e^i +
                  \frac{1}{z-w}\;[J(w),\n(w),e^i\,]_{X(w)}\non
 G^a(z)\,G(w) &\sim& \frac{2}{(z-w)^2}\;e^a\circ_{X(w)}J(w) +
                  \frac{1}{z-w}\;\partial_w(e^a\circ_{X(w)}J(w)) \non
&&{}+\frac{2}{z-w}\,e^aL(w)\nn
\ea
\\
It should be emphasized that the realization (\ref{parant}) of the algebra is
distinguished in a way, because its free fields (\ref{ff}) are intimately
related to the structure ``constants'' via (\ref{conn}). However, it turns out
that the generators of all known free field realizations of the soft algebra
\cite{berk, brcedpr} include exactly one copy of (\ref{parant}), that is
responsible for the correct transformation behavior of the structure
``constants'' under the algebra. We will call this part (\ref{parant}) the {\em
parameter field part} of the generators. The remaining part of the generators
in extended realizations will be referred to as the {\em main part}.\\
\paragraph{Extended realizations}
The main part of the generators found by Brink, Cederwall and Preitschopf
\cite{brcedpr} is built of additional octonionic free fields $S^a$ and $\f^a$,
fermionic and bosonic respectively, with fundamental correlator
\ba
\f^a(z) \f^b(w) &\sim& \d^{ab} \ln|z-w| \label{ffkm}\\
S^a(z) S^b(w) &\sim& \frac{\d^{ab}}{z-w} \nn
\ea
The complete generators are given by
\ba
  J &=& j + \halb S^*\circ_XS \label{n8} \\
  G &=& g + \partial\f^*\circ_XS
                + \halb [S,S,\n]_{X} \non
  L &=& l + \halb:[\partial\f^*\partial\f]:
                - \halb :[S^*\partial S ]: \nn
\ea
and can be shown to fulfill the OPE (\ref{OPEkl}) classically. Here, the
algebra appears in connection with the super-Poincar\'e-algebra of the
ten-dimensional N=1 GS-superstring in light-cone gauge. The physical meaning of
the parameter fields (\ref{ff}) remains unclear. \\
Note that this realization is the generalization of the well known free field
representations of superconformal algebras in superstring models for N=1, N=2
and N=4, that consist of one copy of the main part from (\ref{n8}) for each
(real, complex or quaternionic, respectively) space-time dimension
\cite{rsstring, ademollo2, ademollo4}.\\
\\
Another realization of the soft N=8 superconformal algebra was found by
Berkovits in the context of a string-twistor-model of the ten-dimensional N=1
GS-superstring \cite{berk}. The parameter field part is built of free fields,
that parameterize the redundant degrees of freedom in the twistor description.
The main part of the generators consists of bosonic $(\L^a, \P^a_{(\L)})$ and
fermionic $(\Th^a, \P^a_{(\Th)})$ constraints with fundamental correlations
\ben
\L^a(z)\P^b_{(\L)}(w) ~\sim~ \frac{\d^{ab}}{z-w} ~\sim~ \Th^a(z)\P^b_{(\Th)}(w)
\een
The explicit form of the generators is given by
\ba
  J &=& j + \{\P_{(\L)}^*\circ_X\L\} \label{n8b} \\
  G &=& g + \P_{(\Th)}^*\circ_X\L - \partial\Th^*\circ_X\P_{(\L)} +
[\P_{(\L)},\L,\n]_X \non
  L &=& l + \halb:[\L^*\partial\P_{(\L)} - \partial \L^*\P_{(\L)}]:
           - :[\P_{(\Th)}^*\partial\Th]:~\nn
\ea
and they also fulfill the classical OPE (\ref{OPEkl}).\\
\\
General extended realizations of the N=8 superconformal algebra can be built of
one copy of the parameter field part (\ref{parant}) and arbitrary copies of the
main parts from (\ref{n8}) and (\ref{n8b}).\\
More complicated extended realizations should exist in analogy to the
constructions for the soft $\widehat{S^7}$ algebra corresponding to certain
higher tree-graphs \cite{cedpr}.

\resection{Quantizing the algebra}\label{qf}
We are now going to investigate the quantized form of the OPE (\ref{OPEkl}). In
particular, we will calculate the OPE of the generators (\ref{n8}) considering
also the terms of order $\hbar^2$, which descend from double contractions and
from reordering the non-commuting free fields the generators are built of. In
the case of the well-known superconformal Lie-algebras such terms are always
central extensions of the algebra. In addition to these c-number-anomalies, the
soft N=8 algebra yields also terms containing the parameter-fields $\l$ and
$\th$, which result from derivations of the structure ``constants''. This was
already noticed by Cederwall and Preitschopf in their analysis of the
$\widehat{S^7}$ algebra.
We will calculate all the anomalies proceeding in three steps:
\begin{itemize}
\item{The OPE of the generators (\ref{n8}) are calculated according to Wick's
theorem in free-field normal-ordering. Up to central extensions all the
additional terms descend from double contractions of the generators' main
parts.}
\item{We show how all these anomalies can be annihilated by adding suitable
quantum corrections to the generators of the algebra. In particular, the OPE of
all the realizations described in the previous section can be brought into the
same form with all soft anomalies vanishing.}
\item{The corrected OPE just have the classical form (\ref{OPEkl}) together
with canonical central extensions. However, the structure ``constants'' and the
currents on the r.h.s. of these OPE being operators with possible singularities
in their short-distance behavior now, are free-field normal-ordered. These
expressions might depend on the special choice of a free-field representation
of the algebra, which seems to be unnatural from the abstract algebraic point
of view. The transition to a more natural so-called current normal-ordering
yields additional terms in the OPE, which descend only from the parameter field
part of the generators.}
\end{itemize}
\paragraph{Anomalies from double contractions}
We find it convenient to express the arising anomalies via the following
combinations of parameter fields:
\ba
 \nu       &:=& \{\partial\l^*\l\}\frac{1}{\l\l^*} = \{\partial\l^*\l^{-1*}\}
                 = \partial X^*X = - X^*\partial X
                    \label{defnun}\\
 \hat{\nu} &:=& [\partial\l^*\l]\frac{1}{\l\l^*} = [\partial\l^*\l^{-1*}]
                    = \frac{1}{|\l|}\;[X^*\partial\l] =
\frac{1}{|\l|}\;\partial|\l|\non
 \n        &:=& \partial\th^*\l\frac{1}{\l\l^*} = \partial\th^*\l^{-1*}
                    = \frac{1}{|\l|}\;\partial\th^*X\nn
\ea
\\
A somewhat tedious but straightforward calculation making repeated use of
Wick's theorem and the identities (\ref{oi})--(\ref{gut}) yields the following
exact quantized form of OPE for the generators (\ref{n8}):
\jot3pt
\ba
 J^i(z)\,J^j(w) &\sim& \frac{8-4\s}{(z-w)^2}\;\d^{ij} +
                    \frac{2}{z-w}\;T^{ijk}J^k
                    + \;\s\left(\frac{4}{z-w}\;T^{ijk}\nu^k
\right)\label{OPEs}\\
\non
 J^i(z)\,G^a(w) &\sim& \frac{1}{z-w}\;\D^{bia}G^b +
                    \frac{2}{z-w}\;R^{airj}\n^rJ^j
                     \non
                &+&  \s\left( \frac{4}{(z-w)^2}\;T^{air}\n^r +
                   \frac{4}{z-w}\;\left(R^{aikr}\nu^k\n^r
                                  - 2T^{ikm}T^{mar}\nu^k\n^r\right)
\:\right)\non
&&\non
G^a(z)\,G^b(w) &\sim& \frac{-16+8\s}{(z-w)^3}\;\d^{ab} +
                    \frac{2\D^{aib}J^i}{(z-w)^2} +
                    \frac{1}{z-w}\;\Big(\partial(\D^{aib}J^i) + 2\d^{ab}L\Big)
+ \non
                &+&  \s\left( -\frac{8}{(z-w)^2}\;T^{abk}\nu^k +
                    \frac{4}{(z-w)^2}\;\left(R^{abrs}\n^r\n^s -
                                        2\{\n\}^a\{\n\}^b\right) \right.\non
           & & \hspace{5mm} {}-\frac{4}{z-w}\;\partial
\left(T^{abk}\nu^k\right)
                  + \frac{2}{z-w}\;\partial \left(R^{abrs}\n^r\n^s -
                                              2\{\n\}^a\{\n\}^b \right)\non
           & & \hspace{5mm}{}+ \frac{8}{z-w}\;\Big(2T^{rs(a}\nu^{b)}\n^r\n^s -
T^{rm(a}R^{b)skm}\nu^k\n^r\n^s \Big) \non
& & \hspace{5mm} {}+ \frac{8}{z-w}\;
   \Big(\d^{ab}(1-\d^a)\{\n\}^r\partial\{\n\}^r -
\{\n\}^{(a}\partial\{\n\}^{b)} \Big) \non
           & & \hspace{5mm} {}+ \frac{8}{z-w}\;\Big(\nu^a\nu^b -
\d^{ab}(1-\d^a)\nu^k\nu^k\Big)\bigg) \non
&&\non
L(z)\,L(w)   &\sim&  \frac{-12+6\s}{(z-w)^4} + \frac{2}{(z-w)^2}\;L
                                           + \frac{1}{z-w}\;\partial L, \nn
\ea
\\
with $\s=1$. We will refer to these OPE as {\em $\s$-OPE} and to the
corresponding anomalies as {\em $\s$-anomalies}.\\
\\
A soft N=8 superconformal algebra with generators consisting of the parameter
field part and --- generalizing (\ref{n8}) --- several copies of the main part,
built of octonionic self-conjugate fields $S^\mu$ and $\f^\mu$ with $\mu = 1,
2, \dots, d\,$, fulfills these $\s$-OPE with $\s=d$. In particular, the
parameter field part of the generators itself fulfills the $\s$-OPE with
$\s=0$.\\
The realization of Berkovits (\ref{n8b}), that consists of free fields and
their conjugates, yields the $\s$-OPE with $\s=-2$, as was shown for its
$\widehat{S^7}$ part in \cite{cedpr}.

\paragraph{Quantum correction of the generators}
It is possible to annihilate {\em all} the $\s$-anomalies by adding a suitable
quantum correction to the generators of the algebra.\\
Remember, that the parameter part itself fulfills the OPE (\ref{OPEs}) with
vanishing $\s$-anomalies. We will first construct a correction to this part
that yields the $\s$-OPE with an arbitrary $\s$.\\
\\
The corrected form of $j^i$ was already found in \cite{cedpr}:
\be
j_\s := j + 2\s\nu
\ee
with
\be
j_\s^i(z)\,j_\s^j(w) \sim \frac{8-4\s}{(z-w)^2}\;\d^{ij} +
                    \frac{2}{z-w}\;T^{ijk}j_\s^k
                    + \;\s\left(\frac{4}{z-w}\;T^{ijk}\nu^k \right)
\ee
\\
Inserting this into the desired OPE $j^i_\s g_\s$ from (\ref{OPEs}) yields the
following transformation for the correction of $g$:
\ben
j^i(z)\Big(g_\s(w)-g(w)\Big) ~\sim~ \frac{1}{z-w}\;(g_\s-g)\circ_Xe^i
       +\frac{2\s}{(z-w)^2}\;e^i\circ_X\n +
\frac{2\s}{z-w}\;[e^i,\nu]_X\circ_X\n
\een
A possible correction turns out to be
\ben
g_\s = g - 2\s(\nu\circ_X\n)
\een
Moreover, it can even be checked that every generator of the form
\be
g_\s = g - 2\s(\nu\circ_X\n) + \a(\n\circ_X\nu + \partial\n) + \b\hat{\nu}\n
\label{korrg}
\ee
with arbitrary numbers $\a$ and $\b$ transforms as required.\\
\\
A rather lengthy calculation shows that this correction with $\a=2\s$ and
$\b=0$ yields the right OPE $g_\s^ag_\s^b$~ if in addition $l$ is corrected as
\be
l_\s := l + \s\partial\hat{\nu} \label{korrl}
\ee\\
Making all these corrections hermitian finally, we can summarize our result as
follows:\\
\\
{\em \hspace*{5mm} The generators
\jot4pt
\ba
j_\s(z) &=& j(z) + 2\s\nu(z) \label{korrh}\\
g_\s(z) &=& g(z)-2\s[\nu(z),\n(z)]_{X(z)}+2\s\partial\n(z)
+\frac{\s}{z}\;\n(z)\non
l_\s(z) &=&
l(z)+\s\partial\hat{\nu}(z)+\frac{\s}{z}\;\hat{\nu}(z)+\frac{\s/2}{z^2} \nn
\ea
\hspace*{5mm} fulfill the OPE (\ref{OPEs}).}\\
\\
As all the quantum corrections commute with the main parts of the generators
(\ref{n8}) and (\ref{n8b}), the OPE of any extended realization of the soft N=8
superconformal algebra can be brought into the form (\ref{OPEs}) with an {\em
arbitrary} $\s$ by replacing the parameter field part (\ref{parant}) by the
corrected part (\ref{korrh}). \\
In particular, the $\s$-OPE with $\s=0$ can be achieved, annihilating all the
$\s$-anomalies. The corresponding quantum correction for the generators
(\ref{n8}) for example would be (\ref{korrh}) with $\s=-1$.\\
It seems reasonable to suppose that possible above mentioned higher tree-graph
realizations will yield more-parameter-families of anomalies that may be
arbitrarily shifted by certain corrections, as is the case for $\widehat{S^7}$
\cite{cedpr}.

\paragraph{Normal-ordering of generators and structure ``constants''}
In all the previous calculations the terms have been normal-ordered in
free-field normal-ordering, as is usual in quantum field theory. Two operators
are normal-ordered by decomposing them into the free fields they are built of
and normal-ordering these in a standard way. This is quite practicable as this
free-field normal-ordering is commutative and associative and therefore easy to
handle.\\
However, while working with affine Kac-Moody algebras, another kind of
normal-ordering may be used. The operator fields are taken as generating
functions for their modes, which build the corresponding affine Kac-Moody
algebra. From this point of view the whole information is encoded in the OPE of
these operators. In particular, there is a priori nothing like a
free-field-representation of these operators, that may be a useful but
supplementary tool. Normal-ordering should not depend on a special
free-field-representation of these operators, that may not even be known, but
should be definable just referring to the OPE. \\
A possible definition is the so-called {\em current normal-ordering} \cite{bbs,
fuchs}:
\be
:AB:(z)~ :=~ \frac{1}{2\p i}\oint_z\frac{dw}{w-z}A(w)B(z)\label{nocu}
\ee
This product is neither commutative nor associative, but
\be
:A:BC::(z)-:B:AC::(z)~ =~~::[A,B]:C:(z)\label{no2}
\ee
and
\be
:[A,B]:(z) = \sum_{n=1}^k\frac{(-1)^{n+1}}{n!}\partial^nC^n_{AB}(z) \qquad{\rm
for}\quad A(z)B(w) ~\sim~ \sum_{n=1}^k\frac{C_{AB}^n(w)}{(z-w)^n} \label{no1}
\ee
To distinguish between the two different normal-orderings, we will denote only
the latter by dots:$\quad :AB:$~.\\
In this notation, the free-field normal-ordering of free fields $\phi_1,
\phi_2, \dots, \phi_n$ is given by
\be
\phi_1(z)\phi_2(z)\dots\phi_{n-1}(z)\phi_n(z) ~~=~~
:\phi_1(z):\phi_2(z):\dots:\phi_{n-1}(z)\phi_n(z):\dots::
\ee
Note that also the r.h.s. is invariant under an arbitrary permutation of the
free fields.\\
\\
\\
Returning now to the soft algebra, the OPE (\ref{OPEs}) were calculated with
generators and structure ``constants'' free-field normal-ordered on the r.h.s..
{}From the algebraic point of view this should be expressed now by the current
normal-ordered product just described.\\
Having a closer look at (\ref{OPEs}), the critical terms are seen to result
from the parameter field part only; moreover, they descend only from the terms
containing the conjugate momentum $\p_{(\l)}$, since only these terms have to
be normal-ordered with the $\l$-dependent structure ``constants''. All these
terms are of the form\\
\\
\hspace*{16mm}$::[\p_{(\l)}A^*]:B: $\hspace{27mm} in current normal-ordering \\
\hspace*{6mm}$[\p_{(\l)}A^*]B ~= ~~:[\p_{(\l)}:A^*]B::$\hspace{14.3mm} in
free-field normal-ordering\\
\\
with commuting fields $A^a$ and $B$ (e.g. $A=\l e^{k*}, B=T^{ijk}(X)$ in the
r.h.s. of the OPE $J^iJ^j$). \\
\\
Let us calculate as an example the difference in the case of $B=T^{ijk}(X)$ and
arbitrary $A$. With
\ben
:[\p_{(\l)}A^*]:(z)\quad T^{ijk}(X(w)) ~~\sim~~
\pol\frac{1}{|\l|}\;[(X^*A)[e^j,e^{k*},e^i]_X]
\een
and the help of (\ref{no2}) and (\ref{no1}), the difference between the terms
in different normal-orderings is given by:
\ben
::[\p_{(\l)}A^*]:T^{ijk}(X):-:[\p_{(\l)}:A^*]T^{ijk}(X):: ~~=~~
\frac{1}{|\l|}\;[(X^*\partial A)[e^j,e^{k*},e^i]_X]
\een
\\
All the difference terms can be calculated this way, yielding:
\ba
:j^l:T^{ijk}:: - j^lT^{ijk} &=&2T^{nlm}R^{ijkm}\nu^n +2R^{lijk}\hat{\nu} \\
:j^k:T^{ijk}:: - j^kT^{ijk} &=& 8T^{ijk}\nu^k \label{cp}\non
:j^j:R^{airj}\n^r:: - j^jR^{airj}\n^r &=& 4T^{aim}\D^{krm}(\hat{\nu}+\nu)^k\n^r
- 8R^{aikr}\nu^k\n^r\non
:g^b:T^{abi}:: - g^bT^{abi} &=& -8\,T^{air}\partial\n^r -
8T^{aim}\D^{rkm}(\hat{\nu}+\nu)^k\n^r\nn
\ea
The complete OPE now contain these anomalies and the $\s$-anomalies, that may
be corrected to an arbitrary $\s$, as shown. For simplicity, we state the
complete OPE in current normal-ordering with annihilated $\s$-anomalies:
\jot4pt
\ba
 J^i(z)\,J^j(w) &\sim& \frac{8}{(z-w)^2}\;\d^{ij} +
                    \frac{2}{z-w}\;:T^{ijk}J^k:
                    - \;\frac{16}{z-w}\;T^{ijk}\nu^k \label{OPEn}\\
\non
 J^i(z)\,G^a(w) &\sim& \frac{1}{z-w}\;:(\D^{bia}G^b: +
2:R^{airj}\n^r:J^j::)\non
                &&{}     +
\frac{8}{z-w}\;\Big(T^{air}\partial\n^r+2(R^{aikr}-T^{aim}T^{krm})\nu^k\n^r\Big) \non
G^a(z)\,G^b(w) &\sim& \frac{-16}{(z-w)^3}\;\d^{ab} +
                    \frac{2}{(z-w)^2}\;:\D^{aib}J^i: -
\frac{16}{(z-w)^2}\;T^{akb}\nu^k\non
                &&{}+\frac{1}{z-w}\;\partial(:\D^{aib}J^i:) -
\frac{8}{z-w}\;\partial(T^{akb}\nu^k)
+\frac{2\d^{ab}}{z-w}\,L\non
L(z)\,L(w)   &\sim&  \frac{-12}{(z-w)^4} + \frac{2}{(z-w)^2}\;L
                                           + \frac{1}{z-w}\;\partial L \nn
\ea\\
These anomalies can not be corrected as the previous ones, but, as we shall
soon see, it is exactly this form of anomalies that will be necessary for a
consistent BRST-quantization of the algebra.
\paragraph{Existence of an inverse operator}
There is still one essential point to be mentioned even in the construction of
the classical soft algebra. The whole construction depends heavily on the
existence of an operator
\ben
X(z)=\frac{\l(z)}{|\l(z)|}
\een
parameterizing the seven-sphere. As the existence of inverse fields is highly
nontrivial in conformal field theory, we should spend some comments on the
existence of this operator.\\
Note first, that only the existence of an operator
\ben
Z(z):=|\l^*(z)\l(z)|^{-1}
\een
is required, because all the structure ``constants'' arrive bilinear in $X$ and
the inverse of $\l$ may be expressed as $\l^{-1} =|\l^*\l|^{-1}\,\l^*$.\\
In the previous calculations the OPE of Z have been taken as
\ba
\l^a(z)Z(w) &\sim& 0 \label{ZZ}\\
\p_{(\l)}^a(z)Z(w) &\sim& \frac{2}{z-w}\;\l^aZ^2 \non
Z(z)Z(w) &\sim& 0 \nn
\ea
and nonsingular with all other fields, as is demanded by consistency with
$Z\l^a\l^a=1$.\\
A standard way to obtain such an algebra is now to construct a system of
operators that fulfill (\ref{ZZ}) and divide out the relation $Z\l^a\l^a-1=0$.
This may be achieved from the system of $\l$ and $\p_{(\l)}$ by adding free
fields
\ba
Z(z)\,\p_{(Z)}(w) &\sim& \pol \non
\f(z) \f(w) &\sim& -\ln|z-w| \nn
\ea
and shifting $\p_{(\l)}$ as
\ben
\p_{(\l)}^a \rightarrow \p_{(\l)}^a - 2\p_{(Z)}Z^2\l^a + 4\partial\f Z\l^a
\een\\
It may be verified that these operators yield the desired OPE (\ref{ZZ}) and
that
\be
I=\langle \l^a\l^aZ-1 \rangle \label{ideal}
\ee
is an ideal in the enveloping algebra, that may be divided out.\\
This proves the existence of the algebra. However, we will see that the
existence of this inverse operator will lead to some curiosities in a
Fock-space representation. Considering representations of the algebra on a
Hilbert space $\cal H$, the action of $Z$ is not determined but only restricted
by the relation (\ref{ideal}).\\
Note further, that dividing out the ideal (\ref{ideal}) implies also dividing
out its action on
$\cal H$:~ $\H \rightarrow \H/(I\H)$. To be consistent with the inner product,
$(I\H)$ has to be orthogonal to $\cal H$. This will heavily restrict the
representations. In particular, every state $|\psi\rangle$, that is annihilated
by $\l^a\l^a$, has to be orthogonal to $\cal H$: $\langle\psi|\H\rangle = 0$.
\\
We will further investigate these problems and representations of the algebra
in the next section.\\

\resection{The soft algebra as a gauge algebra}\label{gf}
This section deals with the treatment of the soft algebra arising as a gauge
algebra of constraints. This might happen in some kind of superstring
twistor-model, as is the case in \cite{berk}, or, maybe, on the search for an
N=8 superstring generalizing the known models for N=1, N=2 and N=4.\\
As is usual in gauge theory one would demand the positive modes of the
constraints to annihilate physical states, that are identified by this
condition in the Fock-space of the free fields, the currents are built of.\\
Following this track, we will first construct a Fock-space representation of
the soft algebra, that will yield some unusual features due to the existence of
the inverse operator mentioned above. Finally, we present the BRST operator of
the algebra and show that nilpotency can be achieved by exactly the quantum
corrections of the generators, that were found in the previous section.\\

\subsection{Fock-space representation of the algebra}
The known free-field-realizations of the soft N=8 superconformal algebras
consist of parameter-fields $\l^a$, $\p_{(\l)}^a$, $\th^a$, $\p_{(\th)}^a$ and
fields building the main part of the generators, like $S^a$, $\f^a$, $\L^a$,
$\P_{(\L)}^a$, $\Th^a$, and $\P_{(\Th)}^a$ in (\ref{n8}) and (\ref{n8b}).
Insofar as these fields commute, they can be represented on different
Fock-spaces, the whole Fock-space being obtained by tensoring these
representations. We will not repeat the canonical Fock-space representations of
the latter fields, since they are well known from superstring theory, for
example. Even representations of free fields with ``wrong'' conformal
dimensions, i.e. contradicting spin-statistics, as is the case for $\L^a$,
$\P_{(\L)}^a$, $\Th^a$, $\P_{(\Th)}^a$ and the parameter fields, are known from
the treatment of conformal and superconformal ghosts \cite{fms,lz}.\\
\\
What remains to be done, is to construct a Fock-space representation of the
parameter fields $\l^a$ and $\p_{(\l)}^a$, that is compatible with the
existence of an inverse operator $Z=(\l^a\l^a)^{-1}$. As we shall see, only the
zero modes of these fields will cause some trouble.

\paragraph{Representation of the parameter fields $\l^a$ and $\p_{(\l)}^a$}
It is reasonable to assume an integer moding for these parameter fields:
\ben
\l^a=\sum_{n\in{\bbl Z}} \l^a_nz^{-n-\frac12}
\een
and correspondingly for $\p_{(\l)}^a$. \\
Indeed, from what we saw in the last section it follows:
\be
\l^a\l^a |\W\rangle = (\l^a\l^a)_0 |\W\rangle = \l^a_r\l^a_{-r} |\W\rangle
\not= 0~~, \label{bed}
\ee
because otherwise the vacuum $|\W\rangle$ would be orthogonal to the
Fock-space, implying the vanishing of the inner product. Demanding that all the
positive modes of $\l^a$ should annihilate $|\W\rangle$, (\ref{bed}) implies
the existence of zero modes \footnote{To be exact, it would be enough, of
course, to fulfill this for one $a$ only, giving rise to an integer moding of
$\l^0$ and arbitrary (even twisted) modings of the remaining
$\l^1,\l^2,\dots,\l^7$, for example. This would result in transferring the
representation described above to $\l^0$ only.}.\\
\\
The operators $\l^a_n$ and $\p^a_{(\l)n}$ as well as the modes of the inverse
operator $Z$ can now be represented in a canonical way for $n\not= 0$:
\ben
\l^a_n |\W\rangle = \p^a_{(\l)n} |\W\rangle = Z_n |\W\rangle = 0 \qquad n>0
\een
\hspace{40mm} with\quad $\l^{a\dagger}_n = \l^a_{-n}$, $\p^{a\dagger}_{(\l)n} =
-\p^a_{(\l)-n}$ and $Z_n^\dagger = Z_{-n}$\\
\\
The zero modes remain to be investigated. Non-commutativity
$[\l^a_0,\p^b_{(\l)0}] = \d^{ab}$ and hermiticity $\l^{a\dagger}_0 = \l^a_{0}$,
$\p^{a\dagger}_{(\l)0} = -\p^a_{(\l)0}$ already yield
\ben
\p^a_{(\l)0} |\W\rangle \not= 0\not= \l^a_0|\W\rangle,
\een
showing that it is not possible to represent these operators as multiplication
and derivation, respectively. The same problem arises in the representation of
superconformal ghosts and is treated as follows \cite{lz}:\\
The single Fock-space is replaced by a direct sum of two representations:
\ben
|\W\rangle = |0\rangle + |1\rangle
\een
with $\l^a_0 |1\rangle = \p^a_{(\l)0} |0\rangle = 0$ and the orthogonal
pairing: $\quad \langle0|0\rangle = \langle1|1\rangle = 0$ and
$\langle0|1\rangle = 1$.\\
One might be tempted to repeat this construction or --- more general --- to
consider a direct sum of these representations for $(\l^0_0,\p_{(\l)0}^0),
(\l^1_0,\p_{(\l)0}^1), \dots,(\l^7_0,\p_{(\l)0}^7)$. However, even the
following general ansatz immediately leads to a contradiction:\\
\\
{\em If the vacuum $|\W\rangle$ admits a decomposition
\ben
|\W\rangle = |0\rangle_0 + |1\rangle_0 = |0\rangle_1 + |1\rangle_1 = \dots =
|0\rangle_7 + |1\rangle_7,
\een
with $\l^a_0 |1\rangle_a = \p^a_{(\l)0} |0\rangle_a = 0$ (no sum over $a$),
this will already imply $\langle\W|\W\rangle = 0$.}\\
\\
This follows from (again no summation over $a$ if not explicitly written):
\ba
&& \l^a_0\l^a_0\p^a_{(\l)0} |\W\rangle = \l^a_0\l^a_0\p^a_{(\l)0} |1\rangle_a =
0 \non
&\Rightarrow& \sum_{a=0}^7\l^a_0\l^a_0 \prod_{b=0}^7\p^b_{(\l)0} |\W\rangle = 0
\non
&\Rightarrow& \prod_{b=0}^7\p^b_{(\l)0} |\W\rangle \sim 0 \non
&\Rightarrow& \langle\W|\W\rangle =
\langle\W|\prod_{a=0}^7\l^a_0\prod_{b=0}^7\p^b_{(\l)0} |\W\rangle = 0\nn
\ea\\
A representation of the zero modes has to be of a different kind.
\paragraph{Representation of the zero modes}
We are still looking for a representation of the following system of operators:
\ba
[\l^a_0,\p^b_{(\l)0}]|\W\rangle &=& \d^{ab}|\W\rangle\non
\,[Z_0,\p^a_{(\l)0}]|\W\rangle &=& 2\l^a_0Z_0Z_0|\W\rangle, \nn
\ea
with $\l^{a\dagger}_0 = \l^a_{0}$, $\p^{a\dagger}_{(\l)0} = -\p^a_{(\l)0}$ and
$Z_0^\dagger=Z_0$. \\
\\
Postponing the search for a general representation, let us consider the
constraints on a physical vacuum
\be
J^i_0|\W\rangle = 0 \label{con}
\ee
We will restrict ourselves to the NS-sector of the generators (\ref{n8}) now,
as the main part $\halb(S^*\circ_XS)_0$ of $J_0$ will then automatically
annihilate the vacuum, S having half-integer moding. The whole construction
will become more tedious in the R-sector or in any other realization of the
algebra but will follow the same line.\\
The constraints (\ref{con}) now yield
\ba
\{\l^*\p_{(\l)}\}_0|\W\rangle = 0
&\Rightarrow& \l^*_0\p_{(\l)0}|\W\rangle = [\l^*_0\p_{(\l)0}] |\W\rangle \non
&\Rightarrow& \p_{(\l)0}^a|\W\rangle = Z_0\l^a_0[\l^*_0\p_{(\l)0}] |\W\rangle
\nn
\ea
The action of $\p_{(\l)0}^a$ is thereby determined by the action of the other
operators and of the combination $A:=[\l^*_0\p_{(\l)0}]$. Instead of searching
a representation for $\l^a_0,\p^a_{(\l)0}$ and $Z_0$, it will suffice to
construct a representation for the system $\l^a_0,Z_0,A$ with
\ba
[A,\l^a_0] &=& -\l_0^a \\
\,[A,Z_0] &=& 2Z_0 \nn
\ea
Notice that these commutators imply a grading of the states according to the
eigenvalues of $A$. Note further that $A|\W\rangle$ is again a physical state.
This makes the following assumption possible:\\
\\
{\bf Assumption: \em The vacuum is an eigenstate of $A$:
\be
A|\W\rangle = c|\W\rangle \label{ann}
\ee
\hspace*{25mm}with a real eigenvalue $c \in \bbl R$. }\\
\\
We will first show that this assumption completely determines the whole
representation and the inner product, thereby constructing a consistent
Fock-space representation of the soft algebra. Afterwards we will comment on
its necessity.\\
\\
The representation is now given as follows: The states of the Hilbert-space are
generated from the vacuum by the operators $\l^a_0$ and $Z_0$:
\be
(Z_0)^n(\l^0_0)^{n_0}(\l^1_0)^{n_1}\dots(\l^7_0)^{n_7}|\W\rangle
\ee
dividing out the relation $(Z_0\l^a_0\l^a_0-1)|\W\rangle = 0$.\\
\\
Denote $\O=(Z_0)^n(\l^0_0)^{n_0}(\l^1_0)^{n_1}\dots(\l^7_0)^{n_7}$ and
$|\O\rangle := \O|\W\rangle$, let further $\k_\O$ be the level of this state
with respect to the grading by $A$:
\ben
[A,\O] =: -\k_\O\O,\qquad{\rm i.e.~~} \k_{\l^a_0}=1,~~ \k_{Z_0}=-2
\een
Together with $A^\dagger=-A+8$ this yields:
\be
\langle \O|A\rangle + \langle A|\O\rangle = (8+\k_\O)\langle\W|\O|\W\rangle
\label{grad}
\ee
and inserting $\O=1$ this implies in particular $c=4$; thereby we get the
result:
\be
\k_\O\langle\W|\O|\W\rangle = 0
\ee
This means that the inner product of two states vanishes if they do not have
the opposite level. In particular, only states of level zero have a
non-vanishing norm.\\
\\
It remains to calculate the non-vanishing scalar products:\\
Using $\langle\W|\l^a_0\p_{(\l)0}^b-\p_{(\l)0}^b\l^a_0|\W\rangle = \d^{ab}$ and
$\p_{(\l)0}^a|\W\rangle = 4Z\l^a_0|\W\rangle$, one obtains:
\ben
\langle\W|Z_0\l^a_0\l^b_0|\W\rangle = \frac{1}{8}\d^{ab}
\een
and, extending this procedure:
\ben
\langle\W|Z_0Z_0\l^a_0\l^b_0\l^c_0\l^d_0|\W\rangle =
\frac{1}{80}(\d^{ab}\d^{cd}+\d^{ac}\d^{bd}+\d^{ad}\d^{bc})
\een
In general we arrive at:
\be
\langle\W|Z_0^n\l_0^{i_1}\l_0^{i_2}\dots\l_0^{i_{2n}}|\W\rangle \sim \sum_{\rm
all~pairings}\d^{i_{k1}i_{k2}}\dots\:\d^{i_{k(2n-1)}i_{k(2n)}}~, \label{skal}
\ee
the normalization constant being easily determined from $$\langle\W|\W\rangle =
\langle\W|(\sum_a \l^a_0\l^a_0)^n|\W\rangle$$\\
This finishes the complete description of a Fock-space representation of the
soft N=8 superconformal algebra.\\
\\
We have now given a consistent Fock-space representation of the soft algebra
and even showed its uniqueness under the assumption (\ref{ann}). Let us close
with a few comments on this assumption. Which freedom is left, if (\ref{ann})
is not valid? Of course, only the remaining freedom in the scalar products is
of interest, since states can be arbitrarily identified, if this is consistent
with the scalar product.\\
Since (\ref{grad}) remains correct, the identity
\ben
\langle\O|\p_{(\l)0}^a|\W\rangle = \langle\O|4Z_0\l^a_0|\W\rangle
\een
is still valid for states $|\O\rangle$ of level 1 and it is just in this
context that the identity was used in the calculation of the scalar products
(\ref{skal}).\\
Moreover, $J^i_0|\W\rangle=0$ still implies $\langle\W|\O|\W\rangle = 0$ for
every $\cal O$ with $[J^i_0,\O] \not=0$, i.e. being non-constant on the
seven-sphere.\\
We are left with the unknown scalar products $\langle\W|Z_0^m|\W\rangle$ and
$\langle\W|(\l^a_0\l^a_0)^m|\W\rangle$, that may be assigned arbitrary values
and that determine the real part of all scalar products involving the operator
$A$ via (\ref{grad}).\\
The choice of one of these alternatives should be motivated by physical reasons
in a concrete model. It turns out, that the space of physical states in every
model --- i.e. the space of states annihilated by the positive modes of the
generators of the algebra --- will be very large, for many of the excitations
can be ``made physical'' by a suitable correction in terms of the parameter
fields. This suggests that other restrictions concerning the excitations of
parameter fields in particular, should appear, that might --- among others ---
produce a statement like (\ref{ann}).

\subsection{BRST quantization}
The whole algebra of physical constraints may be encoded in the nilpotent
fermionic BRST-operator with the physical states arising as its cohomology
classes. We have to introduce the corresponding Fadeev-Popov-ghost-fields:
\ba
C(z)B(w) &\sim& \pol \qquad \!{\rm
fermionic~conformal~ghosts~of~dimension~}(-1,2) \non
\c_a(z)\b_b(w) &\sim& \frac{\d_{ab}}{z-w} \qquad {\rm
bosonic~superconformal~ghosts~of~dimension~}(-\halb,{\textstyle \frac{3}{2}})
\non
c_i(z)b_j(w) &\sim&  \frac{\d_{ij}}{z-w} \qquad {\rm fermionic~\widehat{{\it
S}^7}-ghosts~of~dimension~}(0,1) \nn
\ea\\
There exists a standard way to construct the BRST-operator from the OPE of a
(super-)Lie-algebra \cite{ff} \footnote{See e.g. the appendix of \cite{bo} for
the construction in a modern notation.}. The obtained operator is classically
nilpotent, whereas the quantum case has to be investigated separately in
general and yields restrictions on the OPE of the algebra. Since we are dealing
with a soft algebra now, the success of this proceeding is not clear a
priori.\\
However, motivated by the result that the BRST-operator of the soft
$\widehat{S^7}$-algebra exhibits this canonical form and can be made nilpotent
even in the quantum case \cite{cedpr}, we generalize this result to the whole
N=8 superconformal algebra, starting with the following current:
\ba
J_{\rm BRST} &=& c_iJ^i + \c_aG^a + CL - T^{ijk}(X)c_ic_jb_k
             - \D^{aib}(X)\c_ac_i\b_b \\
         &&{}- 2R^{airj}(X)\c_ac_i\n^rb_j -
\D^{ajb}(X)\partial\c_ab_j\c_b-\c_a\c_aB
             + \partial c_ib_iC \non
         &&{}+ \halb \partial C\c_a\b_a - C\partial\c_a\b_a
             + \partial C B C\nn
\ea
or, in octonionic notation:
\ba
J_{\rm BRST} &=& [c^*J] + [\c^*G] + CL - [(c\circ_Xc)b^*] - [(\c\circ_Xc)\b^*]
\hspace{17mm}\\
&& {}- [\,[\c,c,\n]_Xb^*] - [(\partial\c\circ_Xb)\c^*] - [\c^*\c ]B + [\partial
c^*b]C \non
&& {}+ \halb\partial C[\c^*\b] - C[\partial\c^*\b] + \partial C B C\nn
\ea
This current has to be defined only up to total derivatives by which it can be
made a primary field. The BRST-operator is given by:
\be
Q = \frac{1}{2\pi i} \oint dz\; J_{\rm BRST}(z)
\ee
\\
Its classical nilpotency can be shown straightforward. The quantum case
involves quite a lot of additional terms of order $\hbar^2$ resulting from
double contractions and derivations of the structure ``constants'', that make
the calculation rather tedious but yield an interesting structure, as we shall
see. \\
In order to fulfill the nilpotency of the BRST-operator, the vanishing of all
these terms has to be demanded, which determines the exact form of the OPE of
the consistent quantized algebra. In view of the last section we will give the
result in current normal-ordering, thereby independent of any special
free-field representation.\\
The required OPE turn out to be:
\jot3pt
\ba
J^i(z)J^j(w) &\sim& \frac{(8+8)\d^{ij}}{(z-w)^2}+\frac{2}{z-w}\;:T^{ijk}J^k:
-\frac{24}{z-w}\;T^{ijk}\nu^k\label{npa}\\
&&\non
J^i(z)G^a(w) &\sim& -\frac{8}{(z-w)^2}\;T^{air}\n^r + \pol(:\D^{bia}G^b +
2::R^{airj}\n^r:J^j:) \non
&& {}+
\frac{8}{z-w}\;\Big(+4T^{ikm}T^{arm}-2T^{akm}T^{irm}-5R^{aikr}\Big)\nu^k\n^r
\non
&&{}+\frac{8}{z-w}\;T^{air}\partial\n^r \non
&&\non
G^a(z)\,G^b(w) &\sim& \frac{-16-16}{(z-w)^3}\;\d^{ab}+
\frac{2}{(z-w)^2}\;(:J^i\D^{aib}: - 8T^{akb}\nu^k) \non
&&{} +  \frac{1}{z-w}\;\partial(:J^i\D^{aib}: - 8T^{akb}\nu^k) +
                    \frac{2\d^{ab}}{z-w}\,L\non
                & & {}+ \frac{16}{(z-w)^2}\;T^{abk}\nu^k -
                    \frac{8}{(z-w)^2}\;\left(R^{abrs}\n^r\n^s -
                                        2\{\n\}^a\{\n\}^b\right)\non
           & & {} +\frac{8}{z-w}\;\partial \left(T^{abk}\nu^k\right)
                  - \frac{4}{z-w}\;\partial \left(R^{abrs}\n^r\n^s -
                                              2\{\n\}^a\{\n\}^b \right)\non
           &&{}
+\frac{16}{z-w}\;\Big(T^{rm(a}R^{b)skm}\nu^k\n^r\n^s-2T^{rs(a}\nu^{b)}\n^r\n^s\Big) \non
           & & {} - \frac{16}{z-w}\;
   \Big(\d^{ab}(1-\d^a)\{\n\}^r\partial\{\n\}^r -
\{\n\}^{(a}\partial\{\n\}^{b)} \Big) \non
           & & {}- \frac{16}{z-w}\;\Big(\nu^a\nu^b -
\d^{ab}(1-\d^a)\nu^k\nu^k\Big)\bigg)\non
&&\non
L(z)L(w) &\sim& \frac{-12-12}{(z-w)^4}+\frac{2}{(z-w)^2}\;L +\pol\partial L\nn
\ea
\\
Comparing this with the OPE (\ref{OPEs}) and (\ref{OPEn}), we arrive at the
following result:\\
\\
{\em Nilpotency of the BRST operator demands critical OPE of exactly the form
(\ref{OPEn}) with additional $\s$-anomalies (\ref{OPEs}) with $\s=-2$.}\\
\\
As was shown in the previous section, every algebra built from the parameter
field part and arbitrary copies of the main parts from (\ref{n8}) and
(\ref{n8b}) can be made to fulfill these OPE by a suitable quantum correction
(\ref{korrh}) of the generators. In contrast to the known superconformal
Lie-algebras and just as in the case of the $\widehat{S^7}$-algebra
\cite{cedpr}, BRST consistency does not restrict the field contents of the
algebra.\\
The realization of Berkovits \cite{berk} is the simplest realization of the
algebra, that fulfills the critical OPE without any quantum correction. Note,
that the results of Osipov on possible central extensions \cite{osi}, quoted
there, only cover the underlying non-associative Malcev-Kac-Moody-algebras,
thereby neglecting all the soft anomalies.

\resection{Summary}
We have investigated the quantized form of the soft N=8 superconformal algebra
from \cite{berk} and \cite{brcedpr}. The OPE of the known realizations of this
algebra were shown to build a one-parameter-class of OPE in which they can be
arbitrarily shifted by additive hermitian quantum corrections of the generators
(\ref{korrh}). The OPE are given by (\ref{OPEs}) and (\ref{OPEn}) and
parameterized by $\s$.\\
We further found that the OPE, which are necessary for a consistent
BRST-quanti\-zation of the algebra, are a member of this class, thereby
showing, that the BRST-operator of any realization can be made nilpotent by the
corresponding quantum corrections. BRST-cohomology can now be studied in every
concrete model.\\
It seems rather interesting that the same class of anomalies arises in several
contexts here. The possible anomalies are heavily restricted by the
Jacobi-identities, of course, but there are at least several families of them,
as result, for example, from arbitrary shifting the generators by combinations
of the parameter fields (\ref{defnun}).\\
The described Fock-space representation of the soft algebra might serve as a
canonical tool for further investigations, exploring the physical spectrum in
concrete models. In view of the rich structure of this algebra, that is due to
its mathematically exposed position, one should still suppose further
interesting discoveries.
\paragraph{Acknowledgments} This paper grew out of my diploma thesis \cite{dt}
at the IInd Institute of Theoretical Physics at the University of Hamburg. I
would like to thank Prof.\,Dr.\,H.\,Nicolai for leading me to and supervising
this work and Dr.\,R.\,Gebert for many helpful suggestions.\\


\begin{thebibliography}{99}
 \bibitem{ademollo} {\sc M. Ademollo, L. Brink, A. D'Adda, R. D'Auria, E.
Napolitano, S. Sciuto, E. Del Giudice, P. Di Vecchia, S. Ferrara, F. Gliozzi,
R. Musto} and {\sc R. Pettorino}.  {\it Phys. Lett.} {\bf B62} (1976) 105.
 \bibitem{rsstring} {\sc P. Ramond}. {\it Phys. Rev.} {\bf D3} (1971) 2415;\\
                    {\sc A. Neveu} and {\sc J. H. Schwarz}.  {\it Nucl. Phys.}
{\bf B31} (1971) 86.
 \bibitem{ademollo2} {\sc M. Ademollo, L. Brink, A. D'Adda, R. D'Auria, E.
Napolitano, S. Sciuto, E. Del Giudice, P. Di Vecchia, S. Ferrara, F. Gliozzi,
R. Musto, R. Pettorino} and {\sc J. H. Schwarz}. {\it Nucl. Phys.} {\bf B111}
(1976) 77.
 \bibitem{ademollo4} {\sc M. Ademollo, L. Brink, A. D'Adda, R. D'Auria, E.
Napolitano, S. Sciuto, E. Del Giudice, P. Di Vecchia, S. Ferrara, F. Gliozzi,
R. Musto} and {\sc R. Pettorino}. {\it Nucl. Phys.} {\bf B114} (1976) 297.
 \bibitem{class} {\sc P. Ramond} and {\sc J. H. Schwarz}.  {\it Phys. Lett.}
{\bf B64} (1976) 75;\\
{\sc R. Gastmans, A. Sevrin, W. Troost} and {\sc A. Van Proeyen.} {\it Int. J.
Mod. Phys.} {\bf A2} (1987) 195;\\
{\sc Z. Hasiewicz, K. Thielemans} and {\sc W. Troost}. {\it J. Math. Phys}.
{\bf 31} (1990) 744;\\
{\sc F. Defever, Z. Hasiewicz} and {\sc W. Troost}. {\it J. Math. Phys}. {\bf
32} (1991) 2285.
 \bibitem{sohnius} {\sc M. Sohnius}. {\it Z. Phys.} {\bf C18} (1983) 229.
 \bibitem{englert} {\sc F. Englert, A. Sevrin, W. Troost, A. Van Proeyen} and
{\sc Ph. Spindel}. {\it J. Math. Phys}. {\bf 29} (1988) 281.
 \bibitem{berk} {\sc N. Berkovits}.  {\it Nucl. Phys.} {\bf B358} (1991) 169.
 \bibitem{brcedpr} {\sc L. Brink, M. Cederwall} and {\sc C. R. Preitschopf}.
{\it Phys. Lett.} {\bf B311} (1993) 76.
 \bibitem{cedpr} {\sc M. Cederwall} and {\sc C. R. Preitschopf}. {\it Comm.
Math. Phys.} {\bf 167} (1995) 373.
 \bibitem{dixon} {\sc G. Dixon}. preprint hep-th/9410202 (1994).
 \bibitem{fms} {\sc D. Friedan, E. Martinec} and {\sc S. Shenker}. {\it Nucl.
Phys.} {\bf B271} (1986) 93.
 \bibitem{lz} {\sc B. H. Lian} and {\sc G. J. Zuckerman}.  {\it Comm. Math.
Phys.} {\bf 125} (1989) 301.
 \bibitem{ff} {\sc E. S. Fradkin} and {\sc T. E. Fradkina}. {\it Phys. Lett.}
{\bf B72} (1978) 343.
 \bibitem{bo} {\sc F. Bastianelli} and {\sc N. Ohta}. preprint NBI-HE-94-10,
hep-th/9402118 (1994).
 \bibitem{osi} {\sc E. P. Osipov}. {\it Lett. Math. Phys.} {\bf 18} (1989) 35.
 \bibitem{bbs} {\sc F. A. Bais, P. Bouwknegt, M. Surridge} and {\sc K.
Schoutens}. {\it Nucl. Phys.} {\bf B304} (1988) 348.
 \bibitem{fuchs} {\sc J. Fuchs}. {\rm Affine Lie Algebras and Quantum Groups}.
(Cambridge University Press, Cambridge, 1992).
 \bibitem{dt} {\sc J.\,A.\,H.\,Samtleben}. Diploma thesis. (Universit\"at
Hamburg, Hamburg, 1995).
\end{thebibliography}
\end{document}